\documentclass[ aps, 12pt, final, notitlepage, oneside, onecolumn,
nobibnotes, nofootinbib, superscriptaddress, noshowpacs,
centertags]{revtex4}
\usepackage{amsmath}

\begin{document}

\title{Prospects of observing gamma-ray bursts \\ with orbital detectors
	of ultra-high-energy cosmic rays}

\author{\firstname{M.~S.}~\surname{Pshirkov}}
\email{pshirkov@sai.msu.ru}
\affiliation{Sternberg Astronomical Institute, Lomonosov Moscow State University, 119992, Moscow, Russia}%
\affiliation{Institute for Nuclear Research of the Russian Academy of Sciences, 117312, Moscow, Russia}%
\affiliation{P.N. Lebedev Physical Institute, Pushchino Radio Astronomy Observatory,142290, Russia}%

\author{\firstname{M.~Yu.}~\surname{Zotov}}
\email{zotov@eas.sinp.msu.ru }
\affiliation{Skobeltsyn Institute of Nuclear Physics,  Lomonosov Moscow State University, 119991, Moscow, Russia}%

\date{\today}

\begin{abstract}
	TUS (Tracking Ultraviolet Set-up), the first orbital telescope of
	ultra-high energy cosmic rays (UHECRs), has demonstrated that
	instruments of this kind have much broader capabilities and can also
	detect various transient luminous events, meteors, anthropogenic glow
	and other processes taking place in Earth atmosphere in the UV
	frequency range.  In this short letter, we address the question
	whether an orbital detector of UHECRs can also register gamma-ray
	bursts (GRBs) via the fluorescent glow of irradiated nocturnal
	atmosphere.  We analyse the latest Fermi GBM catalog of GRBs and
	properties of several active and perspective instruments.  The study
	reveals that even an advanced detector with parameters of an optical
	system similar to that of the KLYPVE-EUSO (K-EUSO) or POEMMA telescopes
	and an appropriate trigger tuned to register events that evolve much slower
   than an extensive air shower, has very modest
	capabilities in this respect and will be able to observe only a few
	GRBs per year of operation.
\end{abstract}

\maketitle

\section{Introduction}
\label{sec:intro}

Gamma-ray bursts (GRBs) are one of the most violent astrophysical
phenomena: with apparent luminosities that can exceed
$10^{52}~\mathrm{erg~s^{-1}}$, they are observed up to high redshifts
$z=8\text{--}9$.  GRBs comprise two distinct classes of different
origin: ``long'' and ``short'' ones depending on whether their durations
are greater than or less than two
seconds~\cite{1992grbo.book..161K,1992AIPC..265..304D,1993ApJ...413L.101K}.
Long GRBs are thought to be produced during a core collapse of massive
stars, so-called SNIb/c-type supernovae, and have a typical duration
larger than several seconds, while short bursts appear to be associated
with the merger of two neutron stars into a new black hole or a neutron
star with a black hole to form a larger black
hole~\cite{2015PhR...561....1K, GBM:2017lvd}.

The majority of GRBs except for the brightest ones are registered from
space because their $\gamma$-ray spectrum peaks at several hundred keVs,
and Earth atmosphere is absolutely opaque for these photons.  The
attenuation length corresponds to a grammage
$X_0\sim10~\mathrm{g~cm}^{-2}$, which means most of the energy will be
absorbed at altitudes larger than 30~km. Ionization caused by
$\gamma$-rays will produce a considerable fluorescent emission of
nitrogen atoms, which can be registered by a dedicated
instrument~\cite{1997ICRC....7..365C}.  On the other hand, this emission
is similar to fluorescence observed in extended air showers (EAS)
initiated by ultra-high energy cosmic rays (UHECRs) with energies above
$\sim50$~EeV. Thus, orbital detectors of UHECRs
theoretically can register GRB-induced effects as well.
This option attracted considerable interest after TUS, the world's first
orbital telescope of UHECRs demonstrated that a scientific device of this kind
can operate as a multi-purpose instrument capable
of registering ultraviolet (UV) illumination of very different nature.
In particular, it registered hundreds of flashes caused by lightning
strikes, multiple transient luminous events (TLEs), among them
more than two dozen of so called ELVEs
(short-lived optical events that manifest at the lower edge of
the ionosphere as bright rings expanding at the
speed of light up to a maximum radius of $\sim300$~km~\cite{1996GeoRL..23.2157F})
as well as miscellaneous other flashes seemingly related to thunderstorm
activities~\cite{tus-jcap-2017, TUS-TEPA-2016, Zotov:uhecr2016, Zotov:2018, RS}.
It also detected multiple meteors, UV pulsations in the sub-auroral regions,
various anthropogenic lights and
flashes of a yet unknown origin, including a number of unexpected flashes
that demonstrated waveforms and kinematics of the signal very similar to those
expected from EAS but with a much higher luminosity~\cite{poster-fake,Khrenov_2020}.
Besides this, it was demonstrated that even a comparatively simple device like TUS
can register so called nuclearites, nuggets of a hypothetical strange quark
matter~\cite{Kenji-nuggets}.

The surprising variety of different phenomena registered by TUS
motivated us to address the possibility of registering GRBs with this
instrument and a number of other orbital UHECR detectors that are being
actively developed these days.
It was interesting to estimate if an UHECR telescope can
contribute to the GRB science by complementing results of
dedicated experiments.
The point is that none of the existing GRB detectors observes the whole
celestial sphere at any given moment of time.
Thus, there is a chance that a GRB escapes from being registered
just because it took place in a ``blind zone'' of a dedicated instrument.
Registering a GRB with a supplemental tool could be helpful in such a
situation.
It is important to note here that an orbital UHECR detector could
provide information on the time of the trigger, an approximate light curve
of the flash and even an approximate direction to the GRB by an analysis
of non-uniformity of illumination of the field of view of the
detector.\footnote{The latter opportunity was kindly pointed out
	by Toshikazu~Ebisuzaki (RIKEN).}
However, one should take into account
that an observed pattern of GRBs will strongly differ from that of an
EAS (or a TLE): instead of a confined and quasi-linear image of a shower
(or a compact image of a TLE),
the effect will manifest itself in a coherent increase of the
ultraviolet background illumination over the whole field of view of an
instrument since the whole atmosphere will glow brighter for some time.%
\footnote{The TUS telescope did not have side shields, so
    that a trigger could be caused by an illumination of the mirror by
    a flash located far from the field of view, resulting in a signal
    over the whole focal surface. This situation is excluded in the
    future detectors, see below.}
The respective timescale will also be much larger: from a fraction of
a second to possibly a few dozen (and even more) seconds for GRBs
instead of a few dozen microseconds in case of an EAS.

Motivated by the truly multi-purpose nature of TUS,
we employ the Fermi GBM catalog of
GRBs~\cite{2020ApJ...893...46V}
to study if an orbital detector of UHECRs can register a GRB thus
complementing observations of dedicated instruments.
We consider technical parameters of the TUS detector,
the KLYPVE-EUSO (K-EUSO)~\cite{k-euso-ICRC2017}
and POEMMA~\cite{2017ICRC...35..542O} telescopes, which are
being actively developed, and the Mini-EUSO
instrument, which is currently working on board the Russian segment of
the International Space Station (ISS) and is
mostly aimed at studying the UV background of the
night atmosphere of Earth~\cite{mini-Capel-2018}.  The results are
generic and can be easily extended for other detectors.

\section{Orbital detectors of UHECRs}\label{sec:TUS}

The TUS detector was launched into orbit
on April~28, 2016, as a part of the scientific payload of the Lomonosov
satellite and operated till late 2017.
The satellite had a sun-synchronous orbit with an
inclination of $97^\circ\!.3$, a period of approximately 94~min, and a
height of about 500~km.  The telescope consisted of two main components:
a parabolic mirror-concentrator of the Fresnel type and a square-shaped
256-channel photodetector aligned to the focal plane of the
mirror~\cite{tus-ssr-2017,tus-jcap-2017}.  The mirror had an area of
about 2~m$^2$ and a 1.5~m focal distance.  The detector had a
rectangular field of view of $9^\circ\times9^\circ$, which covered an
area of approximately 80~km$\times$80~km at sea level.  The angular
resolution of a single channel was equal to 10~mrad, which results in a
5~km$\times$5~km area at sea level.

The TUS electronics could operate in four modes intended for detecting
various fast optical phenomena in the atmosphere on different timescales.
The main mode was aimed at registering UHECRs and had a time
sampling rate of 0.8~$\mu$s.  Time sampling windows of 25.6~$\mu$s and
0.4~ms were utilized to study TLEs of different
kinds, and a 6.6~ms time bin was available to detect micro-meteors
and possibly space debris.  Each complete data record written in any
mode of operation consisted of 256 time bins of the respective
duration.

The K-EUSO telescope, which is aimed to be installed on board the
International Space Station (ISS) after 2022 for a 2-year mission, is a
much more advanced instrument~\cite{klypve-2015bullras,k-euso-ICRC2015,%
k-euso-ICRC2017,k-euso-ICRC2017-science}.  In its latest version, it is
expected to have a Schmidt-type optical system with the main
mirror-reflector of a 4~m diameter, an entrance pupil of a 2.5~m
diameter and a 1.75~m focal length.  The telescope will have a circular
field of view with a  diameter equal to $40^\circ$, resulting in an
instantaneous geometrical area of nearly $6.7\times10^4$~km$^2$ at sea
level for the ISS altitude of around 400~km.

The focal surface of K-EUSO will have a design similar to that of the
JEM-EUSO telescope~\cite{jemeuso-mission,jemeuso-instrument}.  It will
consist of nearly 120 thousand multi-anode photomultiplier tubes
(MAPMTs) grouped into 52 photodetector modules (PDMs).  A strong point
of the data acquisition system of K-EUSO is its flexibility.  Similar
to JEM-EUSO, the instrument will operate with a sampling time of
2.5~$\mu$s in the main mode of operation, aimed at registering UHECRs.
In case of a trigger, the data will be acquired for at least 320~$\mu$s,
i.e., 128 time bins.  Longer sampling rates can be employed to register
slower phenomena such as TLEs, meteors or nuclearites (strange quark
matter).  For example, a sampling time of 2.56~ms
($=1024\times2.5~\mu$s) with the total record duration of~2.6~s can be
used for registering meteors~\cite{jemeuso-meteors}.

One of the most ambitious orbital projects in the field of UHECRs is the
POEMMA experiment~\cite{2017ICRC...35..542O}.
POEMMA will consist of two identical satellites flying in concordance
at low Earth orbits with the ability to observe overlapping regions of
the nocturnal atmosphere at angles ranging from nadir to just above the
limb of Earth.
The altitudes of the satellites will vary from 525~km up to 1000~km
with different separations and pointing strategies.
Similar to K-EUSO, the optical system of each detector will consist of
a Schmidt-type telescope with a $45^\circ$ field of view and an
optical aperture of 2.3~m in diameter with a single correction plate.
It was estimated that the total area of
observation by two satellites at an altitude of 525~km and separation of
925~km operating in the stereo mode for registering UHECRs will be of
the order of $3.24\times10^5~\mathrm{km}^2$~\cite{2017ICRC...35..542O}.

Finally, let us briefly consider Mini-EUSO, a small UV telescope that
is operating on board the ISS since October, 2019,
looking down on Earth through a nadir-facing, UV-transparent window
from inside the Russian Zvezda module~\cite{mini-Capel-2018}.
Its main scientific goal is to
produce a high-resolution map of UV emissions from Earth.  It can also
register TLEs, meteors, space debris, nuclearites, bioluminescence, thus
obtaining detailed information about the UV background level, necessary
for the successful development of K-EUSO and other experiments of the
EUSO program~\cite{EUSO-program-ICRC-2017}.
Strictly speaking, Mini-EUSO is not an UHECR detector but it is
interesting to look closer at its technical capabilities since it is one
of the pathfinders of the future UHECR missions.

Mini-EUSO is based on a single PDM, which consists of 36 
MAPMTs, each with 64 pixels. An optical system of two
double-sided Fresnel lenses with a diameter of 25~cm is employed to
focus light on the PDM in order to achieve a circular field of view
of~$44^\circ$, which results in $\approx8.2\times10^4~\mathrm{km}^2$ at
sea level.  The aperture of the detector equals 490~cm$^2$.  The PDM
detects UV photons and is read out by the data acquisition system with a
sampling rate of 2.5~$\mu$s and a spatial resolution of 6.11~km.  The
instrument utilizes a multi-level trigger
system~\cite{mini-Capel-2018,mini-Belov-2018}.

The first-level trigger of Mini-EUSO is configured to detect UHECR-like
events with a time resolution $\tau=2.5~\mu$s. An event is triggered if
a signal collected in~20~$\mu$s exceeds the background level
by~$8\sigma$. A whole record consists of 128 time steps thus occupying
320~$\mu$s.  At the second level, the time resolution equals 320~$\mu$s.
Once again, a record consists of 128 samples, giving 40.96~ms of data.
The level of excess of the signal over the background and the number of
time samples used to estimated the signal can be altered during the
flight.  This mode works independently from the first one and is
intended for registering transient luminous events and other phenomena
with a similar duration.
The ``slow''
second-level trigger is implemented in Mini-EUSO for the very first
time in the whole JEM-EUSO family of detectors~\cite{mini-Belov-2018}.
Besides these, a continuous readout with a
resolution of 40.96~ms is implemented. This mode dose not have a trigger.
It is intended for mapping the Earth UV background and to search for
meteors, space debris and strange quark matter during an offline
analysis. Duration of one record in this mode equals 5.24~s.

\section{Sensitivity of an UHECR telescope to GRBs}
\label{sec:sens}

As was already mentioned above, photons coming from a GRB are mostly
absorbed at altitudes higher than 30~km.  Only a minor fraction of their
energy is radiated in the fluorescence process, while the rest is lost
in collisions and internal quenching. 
The key question is the fluorescence yield of photons in air.

The pioneering studies of fluorescence emission in gases induced by
rapid particles date back to mid-1950's~\cite{Grun1954},
see~\cite{Arqueros2008} for a review of this and other early works.  The
first investigation of nitrogen fluorescence emission with respect to
cosmic-ray detection (i.e., for electron--air collisions) was performed
some ten years later, when comprehensive measurements were accomplished
by Bunner~\cite{Bunner1967}.  The 1970's gave rise to an interest in the
possibility of detecting astrophysical X-ray~\cite{Elliot1972} and
gamma-ray transients~\cite{Weekes1976} via atmospheric fluorescence.
It was shown in particular that when X-rays interact with air, their
energy is quickly converted to electron energy via the photoelectric
effect or Compton scattering, so that one can use results for electron
measurements to infer the fluorescence efficiency of air when excited by
X-rays.
The absolute fluorescence efficiency at low pressure was found to be
$\approx0.0035$ for both photon--air and electron--air collisions at
energies from a few keV to 100~keV, with the efficiency for lower energy
photon--air collisions being only slightly lower than that for
electron--air, see Fig.2-4 in~\cite{Elliot1972}.
Later on, Catalano et~al.\ studied if gamma-ray bursts can be registered with a
dedicated orbital detector by the fluorescence emission of the
atmosphere~\cite{1997ICRC....7..365C}. They adopted the fluorescence
efficiency in the range 0.002--0.004 basing on the results by
Bunner~\cite{Bunner1967}.

Since the fluorescence efficiency of photon--air collisions equals that
of electron--air collisions at low density and pressure
($\rho\sim2\times10^{-4}~\mathrm{g~cm}^{-3}$, $p\sim1~\mathrm{kPa}$) in
the energy range of interest, one can employ newer experimental results.
In what follows, we adopt the fluorescence efficiency
$\eta\approx3.5\times10^{-3}$ following~\cite{2003APh....20..293N,
2004APh....22..235N}.
The estimate was obtained using the total efficiency in all bands in the
300--400~nm range and taking into account a low pressure of the
environment, see Table~3 and Eq.~(8) in~\cite{2003APh....20..293N}.  The
number of signal photons from a GRB can be written as
\[
	N_s=\frac{\eta}{4\pi\epsilon}A \Omega F_\mathrm{GRB},
\]
where $\epsilon\sim$3--4~eV is the typical energy of a UV photon in the
range of interest (300--400~nm), $A$ is the area of the optical aperture
of an instrument, $\Omega$ is the size of its field of view, and
$F_\mathrm{GRB}$ is the fluence of a GRB.

The rate of background illumination for a particular orbital instrument
can be written as	$\mathcal{R}_b=\mathcal{B} A \Omega$, where the
background level~$\mathcal{B}$ is the UV glow of the nocturnal
atmosphere. Observations performed with Tatiana and Tatiana-2 satellites
demonstrated that it varies in a broad range
$\mathcal{B}=3\times10^{7}$--$10^{8}~\mathrm{photons}~\mathrm{cm^{-2}~s^{-1}~sr^{-1}}$
even during moonless nights, depending on the type of the surface with
the lowest levels at medium latitudes above deserts, forests and oceans
but also depending on the cloud coverage, and it can be an order of
magnitude higher during full-moon periods~\cite{2005APh....24..400G}.

It is now straightforward to estimate the signal-to-noise ratio
SNR for a particular GRB and an orbital instrument:%
\footnote{In general case, one should include an impact of the signal
in the denominator of the formula for the SNR. In our case,
$N_s\ll \mathcal{R}_b t_\mathrm{GRB}$ so that we can omit this term.}
\[
	\mathrm{SNR} = \frac{N_s}{\sqrt{\mathcal{R}_b t_\mathrm{GRB} }}
		=\frac{\eta}{4\pi\epsilon \sqrt{\mathcal{B}}}\,
		\sqrt{A \Omega} \,
		\frac{F_\mathrm{GRB}}{\sqrt{t_\mathrm{GRB}}},
\]
where $t_\mathrm{GRB}$ is the duration of a GRB.  Assuming the lowest
possible background illumination~$\mathcal{B}$ according to the Tatiana
data, we arrive at
\begin{equation}
	\label{eq:snr}
	\max(\mathrm{SNR}) \approx9.1\times10^3 
		\sqrt{A \Omega} \, \frac{ F_\mathrm{GRB}}{\sqrt{t_\mathrm{GRB}}},
\end{equation}
where $A$ is expressed in units of cm$^2$, $\Omega$ is in steradians,
$F_\mathrm{GRB}$ is in $\mathrm{erg~cm}^{-2}$, and $t_\mathrm{GRB}$ is in
seconds.

We are now ready to estimate capabilities of the detectors discussed
above to register a GRB.  We employed the latest Fermi GBM gamma-ray burst
catalog~\cite{2020ApJ...893...46V} exploiting the high level of
sensitivity of the GBM and its large field of view (8~sr).\footnote{%
	Fermi GBM operates in the 8~keV--40~MeV energy range because prompt
	emission of a ``typical'' GRB peaks at energies $\mathcal{O}$(100)~keV. }
We considered the sample available on May~10, 2020.
The sample contained 2799 events registered in almost 12 years of
operation.  We assumed $t_\mathrm{GRB}=T_{90}$ in equation~(\ref{eq:snr})
and extracted~$T_{90}$ and the fluence~$F_\mathrm{GRB}$ for all
events in the catalog and performed calculations for each of them separately.
We also considered two subsets of the data set:
one with short GRBs with the prompt duration~$T_{90}<2$~s (461 events),
and a complementary set of long GRBs with $T_{90}>2$~s (2338 events).

Finally, one has to estimate corrections coming from the limited
acceptance of an UHECR telescope working as a ``GRB detector.'' First,
the acceptance is constrained by geometric considerations: the zenith
angle~$\theta$ of a potentially detectable GRB must be small enough,
otherwise there will be a considerable suppression: $F_\mathrm{GRB}\to
F_\mathrm{GRB}\cos\theta$.  Assuming conservatively $\theta\sim60^\circ$, we
get~$1/4$ of the celestial sphere.  Next, the fraction of time with the
lowest background level used above is of the order of 15\% of all time
in orbit~\cite{JEM-EUSO-exposure}.  As a result, only $\approx4$\% of
all GRBs can be registered (with an appropriate trigger) by an orbital
telescope of UHECRs.

\begin{table}[!ht]

	\caption{Given in columns: the main parameter of the optical systems
	of the orbital detectors used in Eq.~(\ref{eq:snr}), percentage of
	short (P$_\mathrm{short}$),  long (P$_\mathrm{long}$), and all
	(P$_\mathrm{all}$) GRBs from the Fermi GBM catalog such that
	$\mathrm{SNR}>3$
	for the respective detector.  P$_\mathrm{short}$ and
	P$_\mathrm{long}$ are calculated w.r.t.\ the respective samples,
	P$_\mathrm{all}$ is calculated w.r.t.\ the whole sample.  The last
	column ($\bar N_\mathrm{GRB}$) shows the mean number of GRBs
	expected to be registered by the respective instrument in one year of
	operation.  For POEMMA, the first line is for a single telescope, the
	second one is for the UHECR stereo observation mode.  We take the
	area of the optical aperture of a single telescope but assume that
	the whole observation area available in the stereo mode can be used
	to estimate the SNR.  }

	\label{tab1}
	\medskip
	\centering
	\begin{tabular}{|l|c|c|c|c|c|}
		\hline
		Detector & $\sqrt{A\Omega}$, (cm$^2$~sr)$^{1/2}$ &
		P$_\mathrm{short}$, \% & P$_\mathrm{long}$, \% & P$_\mathrm{all}$, \%
		& $\bar N_\mathrm{GRB}$ \\
		\hline
		Mini-EUSO &   15.0 &      0.4&      1.5&      1.3&      0.2\\
		TUS	    &   22.6 &      1.3&      2.3&      2.1&      0.3\\
		K-EUSO    &  136.4 &      9.5&     17.3&     16.0&      2.1\\
		POEMMA (1)&  141.0 &      9.5&     17.8&     16.5&      2.2\\
		POEMMA (2)&  221.0 &     15.6&     27.2&     25.3&      3.4\\
		\hline
	\end{tabular}
\end{table}

Let us look at the results presented in Table~1.
Fermi GBM has detected 2799 bursts in $\approx11.8$ years of operation with
effective coverage of 2/3 of the celestial sphere.  That means one can
expect K-EUSO to register 2.1 bursts in average every year (with
$\mathrm{SNR}>3$) of
operation in space providing it has an appropriate ``slow'' trigger.
Approximately the same number of registered GRBs can be expected from
any single telescope of the POEMMA mission.
The stereo mode of observing UHECRs extends the capabilities of POEMMA
to register GRBs by approximately~1.5 times.
The numbers will increase up to~3.1 and 4.8 respectively if one takes
$\mathrm{SNR}>2$.
Chances of Mini-EUSO and TUS to detect a GRB are slim.

\section{Conclusions}

It was demonstrated by the TUS experiment that orbital detectors of
ultra-high-energy cosmic rays can be useful is studying a variety of
other phenomena.  In this paper, we addressed the question if such
instruments are able to detect gamma-ray bursts via the fluorescent glow
of irradiated nocturnal atmosphere to complement observations of
dedicated experiments.  Our estimates made for several active and
next-generation projects show that their capabilities are quite modest
in this respect.  Even an advanced detector with parameters of an
optical system similar to that of the K-EUSO or POEMMA telescopes and an
appropriate trigger tuned to register events that evolve much slower
than an extensive air shower, will be able to observe only 2--3 GRBs per
year of operation.

\bigskip
\paragraph*{Acknowledgements.}

We thank the anonymous referee for numerous insightful comments that
allowed us to clarify some important issues in the text.  We also thank
Mario~Bertaina for a valuable comment, and Alexander~Belov and
Pavel~Klimov for helpful discussions of the design of K-EUSO and
Mini-EUSO.  This research has made use of the NASA/IPAC Extragalactic
Database (NED), which is operated by the Jet Propulsion Laboratory,
California Institute of Technology, under contract with the National
Aeronautics and Space Administration.
The authors acknowledge support of the Interdisciplinary Scientific and
Educational School of Moscow University ``Fundamental and Applied Space
Research.''


\end{document}